\documentstyle[pra,aps,twocolumn,epsf]{revtex}
\begin{document}
\draft
\title{The dipole polarizability of the hydrogen molecular ion}
\author{J. M. Taylor, A. Dalgarno, and J. F. Babb}
\address{
Institute for Theoretical Atomic and Molecular Physics,\\
Harvard-Smithsonian Center for Astrophysics,\\
60 Garden Street, Cambridge, MA 02138}

\maketitle
%\newpage
%%%%%%%%%%%%%%%%%%%%%%%%%%%%%%%%%%%%%%%%%%%%%%%%%%%%%%%%%%%%%%%%%%%%%%%
\begin{abstract}
A procedure is described for the precise nonrelativistic evaluation of
the dipole polarizabilities of $\mbox{H}_2{}^+$ and $\mbox{D}_2{}^+$
that avoids any approximation based on the size  of the electron
mass relative to the nucleus mass.  The procedure is constructed so
that sum rules may be used to assess the accuracy of the
calculation. The resulting polarizabilities are consistent with
experiment within the error bars of the measurements and are far more
precise than values obtained by other theoretical methods.
\end{abstract}
% PACS ok JB

%%%%%%%%%%%%%%%%%%%%%%%%%%%%%%%%%%%%%%%%%%%%%%%%%%%%%%%%%%%%%%%%%%%%%%%
\narrowtext 

The separation of nuclear and electronic motion is the underlying
principle of the theory of molecular structure. The theory is
challenged by recent measurements of Jacobson {\em et
al.\/}~\cite{JacFisFeh98} of the electric dipole polarizabilities of
$\mbox{H}_2{}^+$ and $\mbox{D}_2{}^+$ which have a precision beyond
that obtained in the Born-Oppenheimer approximation.  The measurements
stimulated the introduction of
methods~\cite{SheGre98,Cla98,Mos98,BhaDra99} that take into account
the diabatic coupling omitted in the earlier calculations and they led
to polarizabilities that agree with the measured values within the
combined experimental and theoretical uncertainties. We present here
new theoretical predictions of much greater accuracy which in turn
pose a significant challenge to experiment.
The accuracy of our method can be assessed by the use of sum rules
and we predict nonrelativistically the polarizabilities of
$\mbox{H}_2{}^+$ and $\mbox{D}_2{}^+$ to a precision well beyond that
achieved by the experiments.  
The method is general and it should be possible to
apply it to many-electron diatomic molecules.

Separating out the center of mass motion
we may write for the Hamiltonian of $\mbox{H}_2{}^+$ or
$\mbox{D}_2{}^+$ in an electric field ${\bf F} = F \hat{n}$ lying
along the $Z$-axis of the space-fixed frame
\begin{equation}
\label{ham}
H = -\case{1}{2M}\nabla_R^2 
 -\case{1}{2}(1+\case{1}{2M})\nabla_r^2  + V({\bf r},{\bf R}) 
 + (1+\epsilon)F\hat{n}\cdot{\bf r},
\end{equation}
where $\bf R$ is the vector joining the nuclei, $\bf r$ is the
position vector of the electron measured from the midpoint of $\bf R$,
$M$ is the mass of the proton or deuteron, $V({\bf r},{\bf R})$ is the
electrostatic interaction potential and $(1+\epsilon)= [1+
(1+2M)^{-1}]$. 
We use atomic units throughout.
The change in energy of the system for
small values of the applied field is given by $\Delta E =
-\case{1}{2}\alpha_d F^2$, where $\alpha_d$ is the
polarizability. Thus if $\Psi^{(0)} ({\bf r},{\bf R})$ is the
eigenfunction of the unperturbed system with Hamiltonian $H_0$ and
$E_0$ is the eigenvalue, the polarizability can be written
\begin{equation}
\alpha_d = -2  \langle \Psi^{(1)}  | (1+\epsilon)\hat{n}\cdot{\bf r}|
                         \Psi^{(0)} \rangle ,
\end{equation}
where 
\begin{equation}
(H_0 -E_0) \Psi^{(1)}({\bf r},{\bf R}) +  (1+\epsilon)\hat{n}\cdot{\bf r}
    \Psi^{(0)}({\bf r},{\bf R}) = 0.
\end{equation}
Alternatively $\Psi^{(1)}$ can be determined from the stationary value
of the functional
\begin{equation}
{\cal J} = \langle \Psi^{(1)}  |H_0-E_0|\Psi^{(1)}\rangle
  + 2 (1+\epsilon) \langle \Psi^{(1)}|\hat{n}\cdot{\bf r}| \Psi^{(0)} \rangle .
\end{equation}

If we write  $\Psi^{(1)}({\bf r},{\bf R})$ as an expansion over
some chosen basis set $\psi_n ({\bf r},{\bf R})$,
\begin{equation}
\Psi^{(1)}({\bf r},{\bf R}) = \sum_{n=1}^N Q_n \psi_n ({\bf r},{\bf R}) ,
\end{equation}
assumed to diagonalize the unperturbed  Hamiltonian $H_0$
so that $\langle \psi_n|H_0|\psi_{n'}\rangle = E_n\delta_{nn'}$,
the polarizability may be written
\begin{equation}
\label{alpha-sum}
\alpha_d = 2 (1+\epsilon)^2  \sum_{n=0}^N
\frac{| \langle \Psi^{(0)}|\hat{n}\cdot{\bf r}|\psi_n\rangle|^2}{E_n-E_0}.
\end{equation}
This expression for the polarizability is stationary with respect to
first order errors in $\Psi^{(1)}$ and is bounded from below.

The completeness of the set $ \psi_n ({\bf r},{\bf R})$
can be assessed by inspecting other sum rules.
Introduce the oscillator strength
\begin{equation}
f_n = 2[(E_n-E_0)/ (1+ \case{1}{2M})]
  |\langle \Psi^{(0)}|\hat{n}\cdot{\bf r}|\psi_n\rangle|^2
\end{equation}
and define the sum
\begin{equation}
\label{sum}
S(p) = \sum_{n=0}^\infty 
        [ (E_n-E_0)/(1+\case{1}{2M})]^p   f_n 
\end{equation}
so that
\begin{equation}
\label{polariz}
\alpha_d =  (1+ \epsilon)^2 (1+\case{1}{2M})^{-1} S(-2) .
\end{equation}
Then provided the $\psi_n$ form a complete set, 
\begin{equation}
S(-1) = \case{2}{3}\langle \Psi^{(0)}|r^2| \Psi^{(0)}\rangle  
\end{equation}
and 
\begin{equation}
S(0) = 1.
\end{equation}

The eigenfunctions $\Psi^{(0)}({\bf r},{\bf R})$ and
$\psi_n  ({\bf r},{\bf R})$ can be written as sums of products
of nuclear and electronic wave functions of the
form 
\begin{equation}
\psi_s (\Lambda NM ) =
 \left[ \frac{2N+1}{4\pi} \right]^{1/2} D_{M\Lambda}^{N*} (\Theta,\Phi,0)
 \phi_{s\Lambda}({\bf r},R) \chi_{s\Lambda} (R),
\end{equation}
where $(\Theta,\Phi)$ are angles specifying the orientation 
of the internuclear axis in the space-fixed frame,
$N$ is the total angular momentum quantum number, $M$ is
the projection on to the space-fixed $Z$-axis, $\Lambda$ is
the projection  of the electronic angular momentum on to
the internuclear axis and ${\bf D}$ is the rotation matrix~\cite{Zar88}.
For the ground state of $\mbox{H}_2{}^+$ or $\mbox{D}_2{}^+$, $N=M=\Lambda=0$
and the electronic wave function has $\Sigma_g^+$ symmetry.
The perturbed state is a superposition of states with
$N=1, M=0$ and $\Lambda=0$ and $\pm 1$, the electronic
wave functions having $\Sigma_u^+$ and $\Pi_u$ symmetry.

To calculate the matrix elements of the Hamiltonian
and of the electric dipole operator we transform from the
space-fixed frame to the body-fixed frame following standard
procedures~\cite{Zar88,LefFie86}.
The nuclear kinetic energy operator may be 
written 
\begin{equation}
-\frac{\nabla_R^2}{2 M} =
 -\frac{1}{2M R^2}
  \frac{\partial}{\partial R} R^2\frac{\partial}{\partial R}
  + H_{\rm rot} ,
\end{equation}
where $H_{\rm rot}$ is given by 
\begin{eqnarray}
\label{hrot-expand}
H_{\rm rot} & = &  \frac{1}{2MR^2}
  {({\bf N}-{\bf L})^2} \nonumber \\
 & = & \frac{1}{2MR^2} (N^2 + L^2
  - N^- L^+  - N^+ L^-  - 2 \Lambda^2) ,
\end{eqnarray}
in which $\bf L$ is the electronic angular momentum
and $\pm$ indicates angular momentum raising and lowering
operators. These are the
operators that couple
$\Sigma$ and $\Pi$ states.
The electronic wave functions for $\mbox{H}_2{}^+$  and $\mbox{D}_2{}^+$
are separable in prolate spheroidal coordinates 
and we expressed the electronic basis functions
$\phi_{\alpha\Lambda} ({\bf r},R)$ in terms of these.
The corresponding formulas for the matrix elements of $H_{\rm rot}$
are given by Moss and Sadler~\cite{MosSad89}.
A detailed description of the representation of the nuclear and electronic
eigenfunctions and the construction of the unperturbed
eigenfunction $\Psi^{(0)}$ and the basis functions $\psi_n$
together with a discussion of the convergence properties
is given by Taylor {\em et al.\/}~\cite{TayYanDal99}.

The electric dipole operator must also be transformed
to the body-fixed axis. The necessary procedures are described
by Lefebvre-Brion and Field~\cite{LefFie86}.
For matrix elements of 
$\hat{n}\cdot{\bf r}$ connecting $\Sigma_g^+$ states to
$\Sigma_u^+$ states
\begin{eqnarray}
\label{zme}
 \langle N,\Lambda &=&0 | \hat{n}\cdot{\bf r}|N+1,\Lambda=0\rangle 
   \nonumber \\
 & = & [(N+1)/3]^{1/2} \langle \Lambda=0 |z|\Lambda=0\rangle 
\end{eqnarray}
and connecting $\Sigma_g^+$ states to $\Pi_u$ states
\begin{eqnarray}
\label{xme}
\langle N,\Lambda&=&0 |\hat{n}\cdot{\bf r}|N+1,\Lambda=\pm 1\rangle 
  = \mp [(N+2)/3]^{1/2} 
 \nonumber \\
 & \times &
 \langle \Lambda=0 |2^{-1/2}(x\mp iy)|\Lambda=\pm 1\rangle ,
\end{eqnarray}
where ${\bf r} = (x,y,z) $. The calculation of
$\langle 0 |z | 0 \rangle$ and $\langle 0 |x \mp iy |\pm 1 \rangle$ 
in prolate spheroidal coordinates is straightforward.

Calculations of $S(p)$ were carried out with basis sets $\psi_n$
comprised of electronic and vibrational
functions~\cite{TayYanDal99,Basis-note}. The converged values of $S(0)$
and $S(-1)$ obtained using 121 electronic and 11 vibrational basis
functions are given in Table~\ref{sums}.  The convergence of the sum
rules with basis set size is approximately logarithmic.  Errors were
determined for each sum $S(p)$ by finding $A$ and $c$ such that $A e^{-cn}$
is the difference between the values obtained with basis sets of sizes
$n \times n \times n$ and $(n+1) \times (n+1) \times (n+1)$.  The
total error given in Table~\ref{sums} for each entry is
$A\sum_{t=n}^\infty e^{-ct}= Ae^{-cn}[1-e^{-c}]^{-1}$.

\begin{figure}[htbp]
\epsfxsize=.45\textwidth \epsfbox{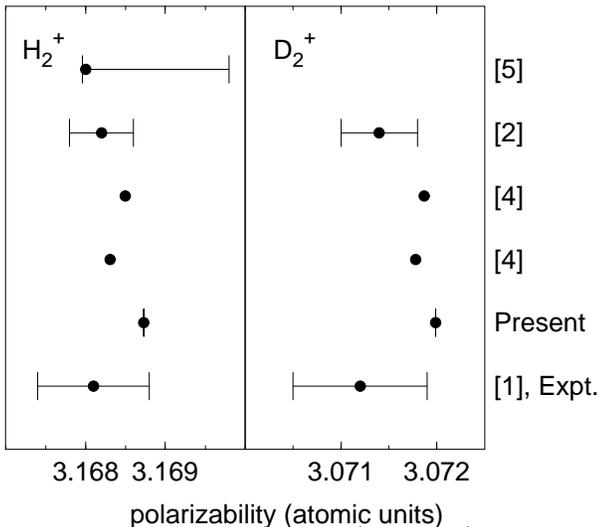}
\caption{Polarizabilities of
$\mbox{H}_2{}^+$ and $\mbox{D}_2{}^+$ in their ground states.
For each of the two calculations
from the present work the error bar is within the vertical
line crossing through the data point.
\label{pol-compare}}
\end{figure}

The values of the calculated sums $S(0)$ and $S(-1)$ agree with the
exact values~\cite{BisChe78b,BabShe92} to better than 2 parts in
$10^8$.  Table~\ref{sums} also lists the values of $S(-2)$, $S(-3)$,
and $S(-4)$.  We anticipate no loss of accuracy in evaluating $S(-2)$
since the summation Eq.~(\ref{alpha-sum}) is stationary with respect
to first order errors.  The corresponding values of the dipole
polarizabilities $\alpha_d$ are given in Table~\ref{pol-table} and
Fig.~\ref{pol-compare}.  The sums $S(-3)$ and $S(-4)$ are related to
quantities occurring in the determination of the
polarizabilities~\cite{JacFisFeh98,StuHesArc88,ArcHesLun90,BabSpr94}.
$S(-3)$ enters in the combination $B_6 \equiv
\case{3}{2}S(-3) - \case{1}{10}C_0$, where $C_0$ is the scalar quadrupole
polarizability. With $C_0= 23.99$ for $\mbox{H}_2{}^+$ and $23.24$ for
$\mbox{D}_2{}^+$~\cite{BisChe78a}, we predict that $B_6= 7.77$ for
$\mbox{H}_2{}^+$ and $7.24$ for $\mbox{D}_2{}^+$.  The empirical value
for $\mbox{H}_2{}^+$ derived by Jacobson {\em et
al.\/}~\cite{JacFisFeh98} is $7.8(5)$.
%%%%%%%%%%%%%%%%%%%%%%%%%%%%%%%%%%%%%%%%%%%%%%%%%%%%%%%%%%%%%%%%%%%%%%%
\begin{table}
\begin{center}
\caption{
Nonrelativistic evaluation of the sum $S(p)$, Eq.~(\protect\ref{sum}), 
for $\mbox{H}_2{}^+$ and $\mbox{D}_2{}^+$.}
\label{sums}
\begin{tabular}{l r@{}l r@{}l}
\multicolumn{1}{c}{$p$} &\multicolumn{2}{c}{$\mbox{H}_2{}^+$} &
   \multicolumn{2}{c}{$\mbox{D}_2{}^+$} \\
\hline
0    &     1&.000\,000\,0(1)   &    1&.000\,000\,0(2)  \\
$-$1 &     1&.653\,650\,96(2)  &    1&.635\,744\,78(6) \\
$-$2 &     3&.167\,000\,94(1)  &    3&.071\,152\,0(2)  \\
$-$3 &     6&.780\,745\,959(7) &    6&.375\,365\,3(3)  \\
$-$4 &  1\,5&.889\,406\,225(5) & 1\,4&.325\,799\,4(6)  \\
\end{tabular}
\end{center}
\end{table}

%\clearpage
%%%%%%%%%%%%%%%%%%%%%%%%%%%%%%%%%%%%%%%%%%%%%%%%%%%%%%%%%%%%%%%%%%%%%%%
\begin{table}
\begin{center}
\caption{
Comparison of theoretical nonadiabatic
values 
of the electric dipole polarizability
for the ground states of  $\mbox{H}_2{}^+$ 
and of $\mbox{D}_2{}^+$ with experimental values.
The results from Refs.~\protect\cite{JacFisFeh98}
and  \protect\cite{Mos98}
have been multiplied by the factor $(1+\epsilon)^2$.}
\label{pol-table}
\begin{tabular}{lll}
\multicolumn{1}{c}{$\mbox{H}_2{}^+$} &
   \multicolumn{1}{c}{$\mbox{D}_2{}^+$} & \multicolumn{1}{l}{Ref.} \\
\hline
 3.168\,0$^{+0.0018}_{-0.0001}$ 
                  & 3.067\,1$^{+0.0016}_{-0.0020}$
                                     &\cite{BhaDra99}, variational\\
 3.168\,2(4)      & 3.071\,4(4)      & \cite{SheGre98}, finite element\\
 3.168\,5         & 3.071\,87        & \cite{Mos98}, artificial channel \\
 3.168\,3         & 3.071\,78        & \cite{Mos98}, variational   \\
 3.168\,725\,6(1) & 3.071\,988\,7(2) &  This work \\
 3.168\,1(7)      & 3.071\,2(7)      & \cite{JacFisFeh98}, experiment\\
\end{tabular}
\end{center}
\end{table}

%%%%%%%%%%%%%%%%%%%%%%%%%%%%%%%%%%%%%%%%%%%%%%%%%%%%%%%%%%%%%%%%%%%%%%%
%\clearpage
%%%%%%%%%%%%%%%%%%%%%%%%%%%%%%%%%%%%%%%%%%%%%%%%%%%%%%%%%%%%%%%%%%%%%%%

Table~\ref{pol-table} and Fig.~\ref{pol-compare} contain a comparison
of our calculated values of $\alpha_d$ with experiment and with the 
results of other theoretical methods. We leave aside calculations of
the polarizability corresponding to an electric field along
the body-fixed axis~\cite{SilBisPip86,Eps87}.
Moss~\cite{Mos98} employed a variational method and an artificial
channel method, with a classical description of the rotation.  We are
able to reproduce his results with our procedure if we take $N=0$ for
the intermediate states with the consequent neglect of
$\Sigma-\Pi$ coupling, the error introduced by ignoring rotational
coupling being accordingly one in the fourth decimal place in the
calculated polarizability.
The calculations of
Bhatia and Drachman~\cite{BhaDra99} and 
Shertzer and Greene~\cite{SheGre98} make no approximations
other than in the numerical applications of their methods
and yield values consistent 
to within the
precision they claim with our results.

We have determined the non-relativistic electric
dipole polarizabilities of the lowest rotational state
of $\mbox{H}_2{}^+$ and $\mbox{D}_2{}^+$ to a precision, we believe, of one
part in $10^8$. 
We expect that relativistic corrections will enter
at the level of one part in $10^5$ based
on known corrections for the hydrogen atom~\cite{ZonManRap72}. 
Other effects arising from the finite size of
the nucleus and nuclear spin will be still smaller.
A new analysis of the experimental data~\cite{JacFisFeh98}
incorporating our values of the sum rules may
yield improved estimates of other properties that enter
the interpretation.

%%%%%%%%%%%%%%%%%%%%%%%%%%%%%%%%%%%%%%%%%%%%%%%%%%%%%%%%%%%%%%%%%%%%%%%
%\acknowledgements
This work was supported in part by the U.S. Department of Energy,
Division of Chemical Sciences, Office of Basic Energy Sciences, Office
of Energy Research.  The Institute for Theoretical Atomic and
Molecular Physics is supported by a grant from the National Science
Foundation to the Smithsonian Institution and Harvard University.

%%%%%%%%%%%%%%%%%%%%%%%%%%%%%%%%%%%%%%%%%%%%%%%%%%%%%%%%%%%%%%%%%%%%%%%
%%%%%%%%%%%%%%%%%%%%%%%%%%%%%%%%%%%%%%%%%%%%%%%%%%%%%%%%%%%%%%%%%%%%%%%

%\bibliographystyle{prsty-noetal}

%%%%%%%%%%%%%%%%%%%%%%%%%%%%%%%%%%%%%%%%%%%%%%%%%%%%%%%%%%%%%%%%%%%%%%%

\begin{thebibliography}{10}
\bibitem{Basis-note}
The basis functions utilized here depend on nonlinear scaling parameters
  $\alpha$, $\beta$, and $\gamma$. The nonlinear parameter $\gamma$ appearing
  in the vibrational sector basis functions of Ref.~\protect\cite{TayYanDal99}
  was replaced in the present work by $\gamma_n = (2n + \beta + 1)^{-1}
  \gamma$, where $n$ is the order of the Laguerre polynomial. Parameters used
  in the present calculations for both $\mbox{H}_2{}^+$ and 
  $\mbox{D}_2{}^+$ are $\alpha = 2.7$,
  $\beta = 77$, and $\gamma = 3500$ for the $\Sigma_u$ functions and $\alpha =
  2.8$, $\beta = 77$, and $\gamma = 3500$ for the $\Pi_u$ functions. They
  differ from those determined in Ref.~\protect\cite{TayYanDal99}, where the
  emphasis was on obtaining minimum values for the energies. In the present
  work we demanded maximum values of the polarizabilities in optimizing the
  $\Sigma_u$ and $\Pi_u$ states. The values of the nonlinear parameters for the
  ground electronic $\Sigma_g$ state are identical with those of
  Ref.~\protect\cite{TayYanDal99}.
\end{thebibliography}
\end{document}